\documentclass[aps,prl,twocolumn,groupedaddress]{revtex4-1}

\usepackage{graphicx}

\usepackage[english,francais]{babel}
\usepackage{amssymb,amsmath}
\usepackage{subcaption}
\begin{document}

\title{Buckling sheets open a door to understanding self-organization in soft matter}

\author{Shankar C. Venkataramani}
\email{shankar@math.arizona.edu}
\affiliation{Department of Mathematics, University of Arizona, Tucson, AZ 85721}







\maketitle

%
%
%
%

Many real world systems are interesting precisely because they can self-organize to form composite structures with novel `emergent' properties -- colloquially, ``More is different" \cite{anderson1972more}. This is certainly true for
living organisms where cells self-organize to form tissues, which make up organs, and so on. It is also true in geological and geophysical systems, materials and composites, and even social structures and
hierarchies.  Thus researchers across many disciplines, grapple with the
following two questions, which are the essence of multiple scale
analysis: (i) how does large scale behavior emerge out of the collective behavior of smaller units?
(ii) what are the rules governing the large scale `system' behavior, and how do these rules influence the behavior of the small scale units? Given the bewildering array of  multiple scale behaviors we see around us, we should not expect  universal answers that apply in all circumstances. In PNAS Davidovitch et al \cite{davidovitch2018geometrically} present some remarkable insights into these questions as they pertain to self-organized structures in thin elastic sheets.  Beyond thin sheets, the ideas in Davidovitch et al \cite{davidovitch2018geometrically} might also apply more generally to {\em energy driven pattern formation} \cite{kohn2007energy}, i.e. to the spontaneous formation of multiple scale structures as a physical system relaxes to its ground state by minimizing its thermodynamic free energy\cite{variational.review}. To discuss the ideas presented by Davidovitch et al  \cite{davidovitch2018geometrically} in PNAS, we first step back for a broader perspective on the subject of self-organization in soft matter.

`Soft matter', a term coined in the 1970's, was initially meant to describe physical systems in which the energy of self-interactions is comparable to thermal fluctutations on a mesoscopic length scale i.e. for a collection  of numerous atoms that is yet much smaller than the macroscopic size of the system \cite{degennes}. Consequently, soft matter is  easily deformed by thermal fluctuations, unlike say, crystalline solids. Concurrently the self-interactions are sufficiently strong that the system can form assemblies that are larger than single atoms or molecules, unlike say gases.  Examples of such materials include colloids, polymers, biological macro-molecules (DNA, proteins, etc.) and liquid crystals.  In current usage, soft matter refers more broadly to physical systems that deform ``easily" under the application of stress. Such systems are governed by  two (or more) distinct physical processes, and the length scale defined by the competition between these processes is mesoscopic.  Examples of self-organization into mesoscopic assemblies are ubiquitous and include diverse phenomena ranging from coagulation of blood to jamming in granular materials to formation of structures  in continuum mechanics (crumpled sheets, blooming flowers, eddies in turbulence, rolls in convection patterns, etc). 

\begin{figure}
\centering
\includegraphics[width=\linewidth]{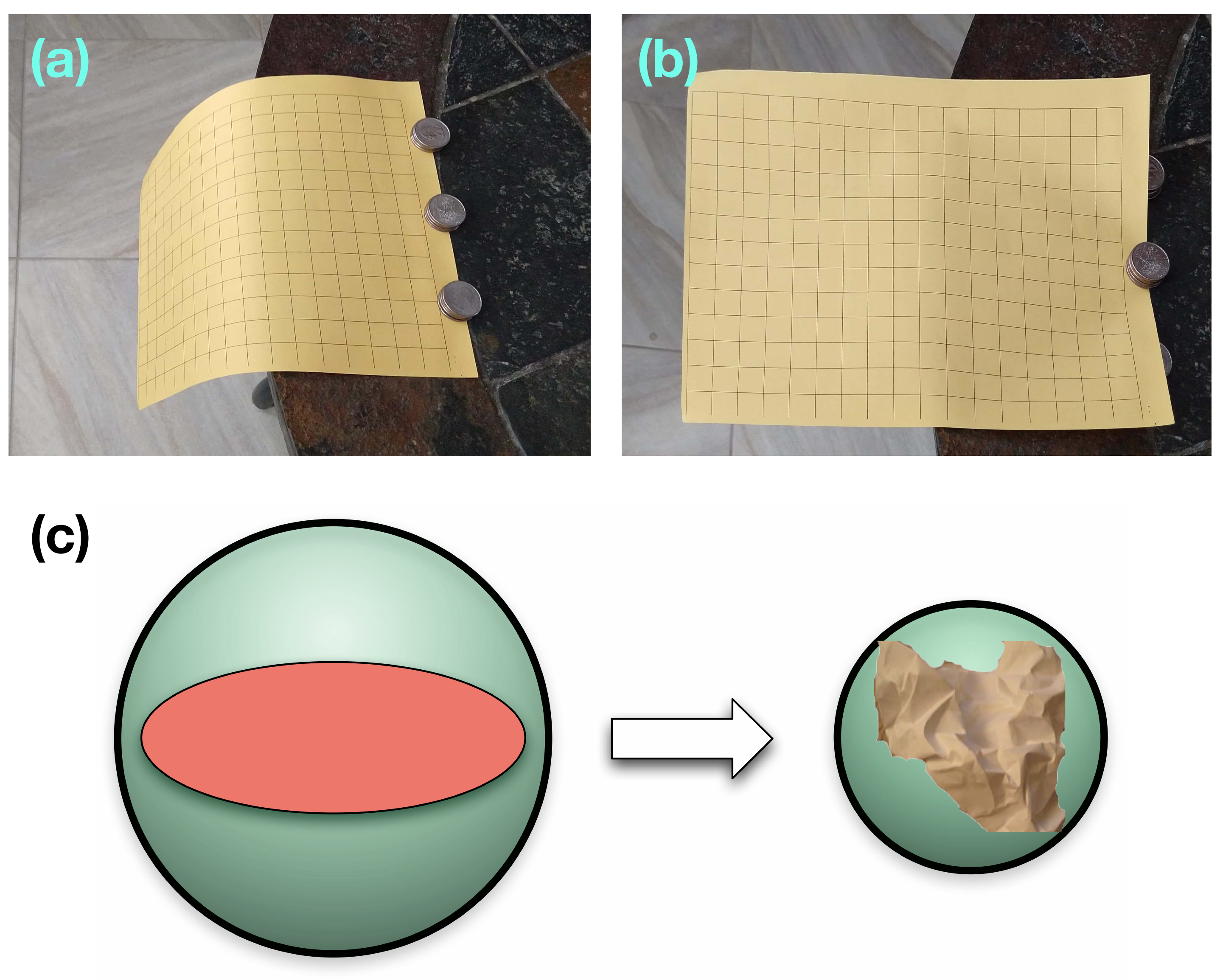}
\caption{(a-b) A letter size paper with a $\frac{1}{2}$ inch grid is held/supported by 3 stacks of 5 quarters, placed 2 inches apart. (a) The sheet is free to bend over the edge of the table without stretching since there is no curvature in the transverse direction. (b) Two of the stacks of quarters are now supporting the sheet from below to introduce a transverse curvature in the sheet. The sheet is now ``rigid" and can no longer bend over the edge without stretching. (c) A thought experiment where a flat sheet if confined within a small sphere causing it to crumple.}
\label{fig:sheet}
\end{figure}

Thin elastic sheets are a quintessential soft-matter system. The ``thinness'' of an elastic sheet is reflected in the
ratio of the flexural and the in-plane rigidities of the sheet, so they are very easy to bend, but much harder to stretch. Furthermore, stretching and bending are not `independent degrees of freedom'. Rather, they are constrained in a non-trivial way by Gauss' Theorema Egregium, which implies that the product of the principal curvatures in an unstretched flat sheet is zero \cite{stoker}. This result is illustrated in figs.~\ref{fig:sheet} (a-b) which capture an effect that is familiar to pizza lovers everywhere.  The physics of thin sheets is driven by the interplay between the low cost of bending $E_b$, the high cost of stretching $E_s$, and  geometric rigidity, i.e.  the ``frustration" of stretching energy inhibiting local bending in two independent directions.

A useful description for multiple-scale behaviors in thin sheets is in terms of localized `defects' and  periodic `microstructure' (respectively `concentration' and `oscillations' in mathematical language \cite{evans}). These structures are emergent features arising from cooperative effects and complicated interactions among many microscopic degrees of freedom  \cite{witten1999insights}. 

An argument for the occurrence of singularities/microstructure
in systems with multiple energy scales goes as follows: 
(i) The weak energy scale involves higher order derivatives
than the strong energy scale (e.g. bending involves the curvature while stretching only involves the deformation gradient); (ii) This results in a singularly perturbed Euler--Lagrange equation; and (iii) Singular perturbation = boundary layers and small scale
structures. 

This chain of arguments explains the crumpling of thin sheets. Gauss' theorem implies that an unstretched sheet has one `locally straight' direction at every point. This precludes the confinement of a thin elastic sheet within a sufficiently small sphere without stretching \cite{immersion_thm},  and a uniformly stretched configuration will be energetically prohbitive. The sheet will thus try to minimize its stretching energy $E_s$ by adopting a non-uniform, multi-scale, crumpled configuration, even when forced in a uniform, large-scale manner. In particular, the energy $E_b + E_s$ in a crumpled sheet condenses on to a network of ridges \cite{science.paper} that meet at point-like vertices \cite{MbAYP97}, and outside these defects, the sheet is essentially stress-free \cite{witten2007stress}. This phenomenon of `stress concentration' is very reminiscent of the non-uniform and highly singular distribution of viscous dissipation in turbulent flows driven by smooth, large-scale stirring  \cite{meneveau1987multifractal}. 

So how is this stress distributed in the sheet? A refinement of the above ideas is to argue that the small scale structures are determined by a balance between the weak energy (i.e. bending $E_b$) and the strong energy (i.e. stretching $E_s$). This idea, motivated by related results in statistical and classical mechanics, goes by various names -- `equipartition', 'dominant balance' and 'virial theorem', and has applications to structure formation in a wide range of disciplines from materials science \cite{maso2006variational} to astrophysics \cite{hansen2004stellar}. Indeed, the key idea for a rigorous analysis of elastic ridges in crumpled sheets is to first show that $E_b E_s^5 \geq O(1)$ for all configurations so that $E_b/E_s \approx 5$ when $E_b+E_s$ is minimized   \cite{Venkataramani_Lower_2004}.
 
The occurrence of singularly perturbed Euler-Lagrange
equations is by itself not a guarantee for the occurrence of multiple
scale behavior in the solutions. An illustrative example \footnote{Tom
Witten showed this to me in 1997}  is
the comparison between a thin elastic rod that is confined
by a ring in two dimensions, and a thin elastic sheet that is confined
by a sphere (See~fig.~\ref{fig:sheet}(c)). The rod ``curls up'' parallel to itself touching the
boundary.  The curvature of the rod is uniform and on the
same scale as the forcing, i.e. the curvature of the confining ring. 
The two dimensional sheet, on the other hand crumples. This example shows that the presence of multiple energy scales, does not in itself create
multiple scale behavior. The geometry of the system plays a
significant role in determining the structures that arise
spontaneously. And, in the case of the rod confined in a ring, $E_s/E_b \to 0$ as the thickness vanishes, so this system does not display equipartition!

The preceding example is often explained away as an exceptional case, where the failure of equipartition is related to the fact that the solutions themselves are not ``multi-scale". Showing remarkable insight, Davidovitch et al \cite{davidovitch2018geometrically} pursue the radical idea that rather than being an exception, this failure of equipartition, $E_s/E_b \to 0$  is actually a widespread feature, that they dub the Gauss-Euler elastica. They ``turn equipartition around" by demanding that $E_s/E_b \to 0$ and use the Gauss-Euler elastica principle as a quantitative tool to calculate multi-scale configurations of thin sheets. In PNAS, Davidovitch et al  \cite{davidovitch2018geometrically} illustrate the applications of their `universal'  Gauss-Euler elastica  by solving two distinct multi-scale buckling problems for thin sheets subject to geometrically incompatible confinements.

Another key idea that is elucidated in this paper as an overarching organizing principle is that, oftentimes, multi-scale phenomena in thin sheets can be driven by two or more small parameters, and one needs to study the relation between the small parameters to deduce the nature of the relevant asymptotic regime (Fig. 4 in Davidovitch et al~\cite{davidovitch2018geometrically} and associated discussion). In particular, the same system can display multiple asymptotic behaviors in different scaling regimes. Generally, one would not expect equipartition, $E_b \approx k E_s$ with $k > 0$, to hold in all of the scaling regimes, and this is indeed a justification for why `universal' principles like the Gauss-Euler elastica will hold in certain asymptotic regimes, while equipartition holds in others.

The ideas in Davidovitch et al~\cite{davidovitch2018geometrically} hold valuable lessons for physicists and mathematicians working on energy driven self-organization. Indeed, the idea that equipartition is not universal is not widely recognized. On the contrary, explanations for multiple scale phenomena based on equipartition, for example the results from~\cite{Audoly_Self} on self-similar buckling in hyperbolic sheets~\cite{eran}, are considered robust and are thus deeply rooted in the physics community even in light of analyses of the phenomenon that lead to lower energy configurations and no equipartition~\cite{EPL_2016}. There is indeed a great need to developing universal principles, akin to the Gauss-Euler elastica in~\cite{davidovitch2018geometrically}, that can be applied to other physical systems in scaling regimes where equipartition does not hold. This work also brings up interesting mathematical questions. Currently, the rigorous mathematical approach to energy driven pattern formation is to recast the problem in terms of (scaling functions of) a single small parameter, say $\epsilon$, and then consider the `limit problem' $\epsilon \to 0$. In this viewpoint, for a problem with multiple small parameters, different scaling regimes correspond to different scaling functions and can lead to different limit functionals \cite{friesecke2006hierarchy}, all of which are reductions of underlying free energy of the system. In PNAS, Davidovitch et al~\cite{davidovitch2018geometrically} consider instead multi-parameter energy functionals, which allows them to investigate the relationships/crossovers between multiple scaling regimes. This will, hopefully, spur mathematical work on formulating multi-parameter models and their limit functionals in a manner that better exhibits the relationships between different asymptotic scaling regimes of the system.

Thin elastic sheets are fascinating. They have many dichotomies and trichotomies -- defects vs microstructure, local vs nonlocal singularities, free vs forced sheets, static vs dynamic phenomena, hyperelastic vs plastic vs growth-induced behavior. Mapping out the phase space of thin sheet behaviors is therefore a challenging enterprise, but the rewards will be well worth the effort, both in terms of theoretical understanding and in terms of practical applications. Key to this program will be  uncovering universal principles that govern these sheets. In PNAS, Davidovitch et al~\cite{davidovitch2018geometrically} have made a significant advance in this direction.

\section*{Acknowledgments}
The author's research is supported by the Simons Foundation through awards 524875 and 560103.


\bibliography{commentary-sv}   

\end{document}